\documentclass[preprint,showpacs,preprintnumbers,amsmath,amssymb]{revtex4}

\usepackage{graphicx}
\usepackage{dcolumn}
\usepackage{bm}
\usepackage{psfig}
\usepackage{epsfig}

\begin{document}

\newcommand{\beq}{\begin{equation}}
\newcommand{\eeq}[1]{\label{#1} \end{equation}} 
\newcommand{\half}{{\textstyle \frac{1}{2}}} 
\newcommand{\gton}{\mathrel{\lower.9ex \hbox{$\stackrel{\displaystyle
>}{\sim}$}}} 
\newcommand{\lton}{\mathrel{\lower.9ex \hbox{$\stackrel{\displaystyle
<}{\sim}$}}} 
\newcommand{\ee}{\end{equation}}
\newcommand{\beqar}{\begin{eqnarray}} 
\newcommand{\eeqar}[1]{\label{#1}\end{eqnarray}}

\title{INITIAL STATE PARTON BROADENING \\ AND  
ENERGY LOSS PROBED IN $d+Au$ AT RHIC}

\author{Ivan Vitev}
 \email{ivitev@iastate.edu}

\affiliation{Iowa State University, Department of Physics and Astronomy 
\\ \hspace*{3cm}  Ames, IA 50010, USA\hspace*{3cm}  }%

\date{\today}

\begin{abstract}

The impact parameter and rapidity dependence of the Cronin effect 
for massless pions in $d+Au$ reactions at  
$\sqrt{s}_{NN}=200$~GeV at RHIC is computed in the framework of pQCD 
multiple elastic scattering on a nuclear target. We introduce a 
formalism to incorporate initial state energy loss in perturbative 
calculations and take into account the elastic energy loss  
in addition to the transverse momentum broadening of partons.
We argue that the centrality  dependence of the Cronin effect can 
distinguish between different hadron production scenarios at RHIC. 
Its  magnitude  and rapidity dependence are shown to carry important 
experimental information about the properties of  cold  nuclear matter 
up to the moderate- and large-$x$ antishadowing/EMC regions.    
 
\end{abstract}

\pacs{12.38.Mh; 24.85.+p; 25.75.-q}

\maketitle

\section{Introduction}

The discovery of jet quenching~\cite{Adcox:2001jp,Levai:2001dc} 
at RHIC and  its  derivative signatures - the high-$p_T$ azimuthal
asymmetry~\cite{Adler:2002ct} and the disappearance  of  back-to-back  
correlations~\cite{Adler:2002tq}  - are among the most direct evidence 
for the creation of dense QCD matter in nucleus-nucleus collisions.   
Recent theoretical work~\cite{Vitev:2002pf,Gyulassy:2000gk} has 
provided successful  qualitative and quantitative explanation of  
the experimental data  via the techniques of perturbative QCD with strong
partonic  final state  interactions leading to non-Abelian energy 
loss and related  modifications to the fragmentation 
functions~\cite{Zakharov:1996fv,Gyulassy:2000fs,Gyulassy:2001nm,Wang:2001if}. 
The important question of the global $p_T$ and $\sqrt{s}$  systematics
of the nuclear modification factors  $R_{BA}(p_T)$ in $p+A$ and $A+A$  
reactions has also been addressed~\cite{Vitev:2002pf}.

Alternative mechanisms related to hadronic rescattering/absorption, 
initial state wavefunction effects, and coherent nucleon scattering  
have been suggested~\cite{Gallmeister:2002us} as possible explanations of 
the observed high-$p_T$ hadron deficit. This letter extends the discussion 
of~\cite{Vitev:2002pf} in light of the upcoming $d+Au$ data and 
demonstrates how the centrality dependence of  the nuclear 
modification factor can distinguish  between contrasting physical 
pictures.  We emphasize the need for comparison between data and theory   
at moderate and high $p_T$  since vastly different 
models give comparable results~\cite{Lin:2003ah} at the level the 
inclusive $dN^{ch}/dy$.

The importance of the $d+Au$ measurements at RHIC, however, reaches 
far beyond  testing the validity of the 
perturbative baseline for jet quenching calculations. In the 
transverse momentum  $2 \leq p_T  \leq 10$~GeV and rapidity 
$-3 \leq y \leq +3$ intervals considered here  the measured hadrons 
retain a memory of the nuclear modifications~\cite{Arneodo:1992wf}  
to the  parton distribution  functions (PDFs) in the  
$10^{-3} \leq  x  \lton  1$ range.  
We find that in the  small-$x$ region of the   nucleus 
the multiple elastic scatterings of the incoming partons completely 
dominate over  nuclear shadowing.  For large nuclear $x$-values  
the EMC effect and the initial state energy loss may  
lead to moderate (20-30\%)  suppression relative to the binary 
collision scaled $p+p$ result over a wide $p_T$ range. The detailed
transverse momentum dependence of the nuclear modification ratio 
may provide valuable complementary information about the gluon 
antishadowing/EMC region, which is not well constrained by data.  
In fact, the  $x$-dependence in the gluon shadowing function  
$S_{g/A}(x,Q^2)$  may be significantly weaker than currently 
anticipated.

\section{ Scaling of hadron spectra in $d+Au$  reactions}

The Cronin effect~\cite{Cronin:zm} is defined as the deviation 
of the total invariant hadron cross section in $p+A$ reactions 
from the binary collision scaled $p+p$ result. 
For the more general case of $B+A$  reactions the 
scaling factor is given by $B \cdot A$. 
The original parametrization of dynamical nuclear effects 
via the Cronin power $\alpha(p_T)$ relates to the   
nuclear modification ratio $R_{BA}(p_T)$ as follows: $R_{pA} (p_T) = 
A^{\alpha(p_T)-1}$. The same relation (without additional factors 
of 2  since the partons from the nucleus do not scatter significantly 
on the deuteron) can also be used to extract a total inclusive 
$\alpha_{dAu}(p_T)$ and a centrality dependent $\alpha_{dAu}(p_T,b)$  
from
\begin{equation}         
R_{d Au}(p_T)  = \left\{   
\begin{array}{ll}  \displaystyle 
\frac{d \sigma^{dAu}}{dyd^2{\bf p}_T}  / 
\frac{A_d \cdot A_{Au} \; d \sigma^{pp}} {dyd^2{\bf p}_T}\, , \;\; &  
{\rm  total \; invariant \; cross \; section  ,}  
\; \;  d+Au  \\[2.8ex]  \displaystyle 
\frac{d N^{dAu}(b)}{dyd^2{\bf p}_T}   / 
\frac{  T_{dAu}(b)\; d \sigma^{pp}}{dyd^2{\bf p}_T}\, , \;\;  &
{\rm about \; impact \; parameter \;  } b,\;  \; d+Au  \; 
\end{array} \right. .  
\label{geomfact} 
\end{equation}          
In Eq.(\ref{geomfact}) $T_{dAu}(b) = \int d^2{\bf r} \; T_{d}({\bf r})
T_{Au}({\bf b-r})$ in terms of the nuclear thickness functions
$T_{A}(b)= \int dz \;\rho_{A}(b,z)$ and the scaling factors are 
computed according to the Glauber model~\cite{Glauber:1970jm}.  
It is important to note that both enhancement ($p_T \gton 1.5-2$~GeV) 
and suppression  ($p_T \lton 1-1.5$~GeV) are an integral part of the 
Cronin effect. Experimentally~\cite{Cronin:zm},  at  $p_T=0.77$~GeV
$\alpha(p_T)$ is below one, leading to a factor $\sim 2$ suppression. 
However, the deviation of $\alpha(p_T)$ from unity is reduced with 
increasing $\sqrt{s}$ and decreasing $x_T = 2 p_T/\sqrt{s}$  - a trend 
exactly opposite to the expectation from strong shadowing and/or gluon 
saturation. The $p_T$ positions of the Cronin peak $R_{pA \; \max}$  
and intercept $R_{pA}=1$ are also stable, i.e. they are
strongly $x_T$-dependent.

The  lowest order pQCD differential cross section for  inclusive 
$A+B\rightarrow h+X$ production that enters Eq.(\ref{geomfact})
is given by~\cite{Vitev:2002pf, Accardi:2002ik}
\begin{eqnarray} 
\left.   
\begin{array}{r}     \displaystyle   \frac{1}{A_d \cdot A_{Au}}  
\frac{d\sigma^{dAu}}{dyd^2{\bf p}_T} \\[2.8ex]  
 \displaystyle  \frac{1}{ T_{dAu}(b)} 
\frac{d N^{dAu}(b)}{dyd^2{\bf p}_T} 
\end{array}  \right\}   & = &  
K  \cdot  \sum_{abcd} \int  dx_a  
dx_b   \int   d^2 {\bf{k}}_a d^2{\bf{k}}_b \,   
\otimes g({\bf{k}}_a) g({\bf{k}}_b)  \otimes  
S_d(x_a,Q^2_a) S_{Au}(x_b,Q^2_b)  \nonumber \\[-1ex] 
&\;& \otimes \; f_{a/d}(x_a,Q^2_a) f_{b/Au}(x_b,Q^2_b)  
\otimes  \frac{d\sigma}{d{\hat t}}^{ab\rightarrow cd} 
\otimes \frac{D_{h/c}(z_c,{Q}_c^2)}{\pi z_c} \; , 
\label{hcrossec} 
\end{eqnarray} 
where we have used the same sets of lowest order (LO)  PDFs, 
fragmentation functions (FFs), and shadowing 
functions~\cite{Eskola:1998df}  as in~\cite{Vitev:2002pf}.
In Eq.(\ref{hcrossec}) the renormalization, factorization, and 
fragmentation scales are chosen to be  $\mu=Q_{a,b}=Q_c=p_T$.

The the $k_T$-broadening function in~(\ref{hcrossec}) approximates 
generalized parton distributions:
\beq
f_{a/p}({\bf k}_T;x_a,Q^2_a) \simeq g({\bf k}_T) \otimes 
f_{a/p}(x_a,Q^2_a), \qquad  
g({\bf k}_T)=  \frac{1}{ \pi \langle {\bf k}_T^2 \rangle_{pp} }  
 \exp \left(- \frac{ {\bf k}_T^2 }
{ \langle {\bf k}_T^2 \rangle_{pp} } \right) \;\;.  
\eeq{gpdf}  
For illustration we include a discussion of parton broadening  
relative to  the axis of propagation in  the  framework of the 
leading  double log approximation (LDLA). The ${\bf k}_T$ probability 
distribution  resulting from vacuum  radiation is given 
by~\cite{Sudakov:1954sw} 
\beq
\frac{1}{\sigma_0} \frac{d\sigma}{d {\bf k}_T^2} {\Big |}_{LDLA} 
= - 2 \,  \frac{ C_R \alpha_s}{2 \pi} \, \frac{1}{ {\bf k}_T^2 } 
\log  \frac{ {\bf k}_T^2 }{Q^2}  \, \exp \left(  
 -   \frac{ C_R \alpha_s}{2 \pi} \,  \log^2 \frac{ {\bf k}_T^2 }{Q^2} 
  \right) \; \; 
\eeq{sudak}   
with  mean  broadening
$ \langle {\bf k}_T^2 \rangle_{pp} = C_R \alpha_s / \pi \, 
(1+ {\cal O}(\alpha_s) )  \, Q^2 $.  $C_R=\{ C_F,C_A \}$ is the 
SU(3) Casimir in the corresponding fundamental or adjoint 
representation. Despite the strong 
momentum  ordering  approximation~\cite{Sudakov:1954sw}, which 
gives an unrealistic null probability for retaining  
the original jet direction,   Eq.(\ref{sudak})  qualitatively  
describes   many the jet physics features. For example, with   
$\half \theta_m = \sqrt{\langle {\bf k}_T^2 \rangle_{pp} / Q^2 } 
\simeq \sqrt { C_R \alpha_s / \pi }$ being the estimate for the 
half-width of the jet cone a slow logarithmic narrowing with $E_T$ 
is predicted. Similarly, to lowest order the cone of a gluon initiated 
jet is found to be 50\% wider ($\sqrt{C_A /C_F}$) than the cone 
of a quark jet.  
  
\begin{table}[htb!]
\begin{center} 
\begin{tabular}{|c||c|c|c|c|} \hline
  $\sqrt{s}$ [GeV]   &  19.4   &  27.4  &  38.8 &  200   \\ \hline 
 $F^h_q(\sqrt{s})$    &  0.220   &   0.150  & 0.096  &  0.015  \\  \hline
\ \ \    $ \langle {\bf k}_T^2 \rangle_{pp}$ [GeV]  \ \ \  & 
 \ \ \ \  1.700   \ \ \ \  & \ \ \  1.775   \ \ \  &  
\ \ \ 1.835 \ \ \ &  \ \  1.920  \ \     \\  \hline
\ \ \     $\Delta(\sqrt{s})$   \ \ \ & \ \ \  \  0.958   \ \ \ \  & 
\ \ \  1.000   \ \ \  & \ \ \  1.034   \ \ \ &  \ \  1.082  \ \    \\  
\hline 
\end{tabular} 
\end{center}
\caption{ The fractional contribution of valance quark hard 
scattering  to hadron production at $y=0,\, \bar{Q}=p_T \simeq 2.5$~GeV 
versus $\sqrt{s}$. The vacuum radiation induced parton broadening 
computed from  Eq.(\ref{ktpp}) and its variation 
$\Delta(\sqrt{s})$ relative to the $\sqrt{s}=27.4$~GeV reference 
are also shown. }
\label{table} 
\end{table}

In this letter we go beyond the  $\sqrt{s}$-independent description 
of the radiative and elastic broadening and take into account the 
initial parton subprocesses for produced hadrons. An increased 
contribution of hard gluon scattering results in an enhanced 
\beq
\langle {\bf k}_T^2 \rangle_{pp}  =  
\left(  C_F F^h_q(\sqrt{s},\bar{Q}) + 
  C_A F^h_g(\sqrt{s},\bar{Q})  \right)  C^2  \;.  
\eeq{ktpp} 
In Eq.(\ref{ktpp})  $F^h_q(\sqrt{s},\bar{Q}), 
F^h_g(\sqrt{s},\bar{Q})=1 - F^h_q(\sqrt{s},\bar{Q}) $ 
are the fractions of hadrons coming from the hard scattering of  
valance quarks and gluons/sea quarks.  With sea quarks we associate 
the Casimir of their  gluon parent ($g \rightarrow \bar{q}q$).  For
hadrons with  $p_T =\bar{Q} \simeq 2-3$~GeV where the $k_T$-smearing 
is reflected, the fraction coming from valence quarks is    
computed from collinear factorized pQCD and given in Table~\ref{table}. 
The corresponding $\langle {\bf k}_T^2 \rangle_{pp}$ is found 
form Eq.(\ref{ktpp}) where $C^2=0.646$~GeV$^2$ has been fixed through  
comparison to the  $\sqrt{s}=27.4$~GeV data~\cite{Cronin:zm} on 
$\half(\pi^+ + \pi^-)$ production in $p+p$. 
$\Delta(\sqrt{s}) = 
\langle {\bf k}_T^2 (\sqrt{s})\rangle_{pp} /   
\langle {\bf k}_T^2 (27.4\,{\rm GeV})\rangle_{pp}$ in Table~\ref{table} 
directly reflects the variation with $\sqrt{s}$ of the effective mean 
color charge of  $\sim$few GeV  projectile  partons relative to the 
$\sqrt{s}=27.4$~GeV case.

We have checked that the energy and transverse momentum dependence 
of inclusive hadron production cross sections is well described with 
the computed values  of $\langle {\bf k}_T^2 \rangle_{pp}$ which 
exhibit a small $< 10\%$ variation about  $\langle {\bf k}_T^2 \rangle_{pp}
 = 1.8$~GeV$^2$ used in~\cite{Vitev:2002pf}.

\section{Implementation of initial state energy loss}

While there are indications that the effect of initial state gluon 
bremsstrahlung may be small~\cite{Arleo:2002ph} 
the elastic energy loss of propagating  hard probes in cold nuclear 
matter has not been considered before. Discussion of QCD scattering 
in nuclear targets can be found in~\cite{Gyulassy:2002yv}. In both 
frameworks of the Qiu-Sterman approach and  the Gyulassy-Levai-Vitev 
reaction operator  approach the {\em longitudinal} correction 
that corresponds to the transverse  momentum broadening can be 
expressed  through the momentum shift operators 
$e^{ - q_{i\,\parallel}\partial_{k_\parallel} } $. The initial parton 
flux  ${d^3N^{i}(k_\parallel,{\bf k}_\perp)}$  flux is propagated      
to the hard collision vertex  according to~\cite{VQ}: 
\beqar
\frac{d^3N^{f}(k_\parallel,{\bf k}_\perp)}{d k_\parallel 
d^2 {\bf k}_\perp }
& = &  \sum_{n=0}^{\infty} \frac{\chi^n}{n!} 
\int  \prod_{i=1}^n   d^2 {\bf q}_{i\,\perp} 
 \left[  \frac{1}{\sigma_{el}} 
\frac{d\sigma_{el} }{d^2{\bf q}_{i\,\perp}} \,
 \left(   e^{-{\bf q}_{i\,\perp} \cdot 
\stackrel{\rightarrow}{\nabla}_{{\bf k}\;\perp }} \otimes 
 e^{- q_{i\,\parallel}\partial_{k_\parallel} }
 - 1  \right) \, \right] \times  \nonumber \\[1ex]   \;  
&& \hspace*{3.3in}\times \, \frac{d^3N^{i}(k_\parallel,{\bf k}_\perp)}
{d k_\parallel d^2 {\bf k}_\perp } \;\;.   
\eeqar{ropit} 
From the twist expansion point of  view the series Eq.(\ref{ropit})
includes the dominant part of all-twist $T=2+2n$   contributions    
$\langle \, q\,  (2n\, FF  ) \, q  \rangle, \langle \, F \, (2n\, FF ) 
\, F \, \rangle $.  Explicit resummation  technique  for (\ref{ropit})  
was first given in~\cite{Gyulassy:2002yv}. 
The interpretation of  our result is simple: 
it folds the Poisson distribution of multiple elastic scatterings 
with  mean  opacity $\chi= \langle L \rangle /\lambda=\bar{n}$,  
which depends on the 
quark/gluon composition of the target and the projectile,
with  the  probability  for transverse deflection per hit given by the 
normalized  elastic  scattering cross section  $ ({1}/{\sigma_{el}}) \,
 {d\sigma_{el}}/{d^2{\bf q}_{i\,\perp}}$. For proton-nucleus 
collisions using a  sharp  sphere approximation  
($T_A({b}) = 3 A \sqrt{R_A^2 - b^2}/(2 \pi R_A^3) $)  
 and assuming equal  scattering probability along the incident 
parton trajectory 
\beq
\langle L \rangle  = \frac{\int d^2 {\bf b} \;  
\sqrt{R_A^2 - b^2}  
T_A ({ b})  }{ \int d^2 {\bf b} \;  
T_A ({ b} )   }  = \frac{3}{4} R_A \;\; ,
\eeq{Leff}  
where we take $R_A=1.2 A^{1/3}$~fm for the nuclear radius.   
One of the important consequences~\cite{VQ} of Eq.(\ref{ropit}) 
is the longitudinal momentum backward shift 
\beq
-{ \Delta k_\parallel} =  {\mu^2 \chi  \xi} \;
 \frac{1}{2 k_\parallel} \;  
\eeq{parshift}
that couples to the mean  $k_\perp$-broadening  
$ \langle \, \Delta k_\perp^2 \, \rangle_{pA} = {\mu^2 \chi  \xi}$. 
For the Gaussian approximation to multiple elastic scatterings, 
$\xi = const. = {\cal O}(1)$.  Beyond this analytically convenient 
ansatz the power law tails of the distribution of the scattered partons, 
resulting from harder fluctuations along the projectile path, lead 
to a logarithmic enhancement of the  mean $k_\perp$-broadening. This 
result is the analogue of the corrections to  the Moliere multiple
collision series in QED. We use  $\xi = \ln(1 + \tilde{c} p_\perp^2)$ 
as in Ref.~\cite{Vitev:2002pf} with $\tilde{c} \simeq 0.14$~GeV$^{-2}$.

\begin{figure}[t!] 
\vspace*{0.5cm }
\epsfig{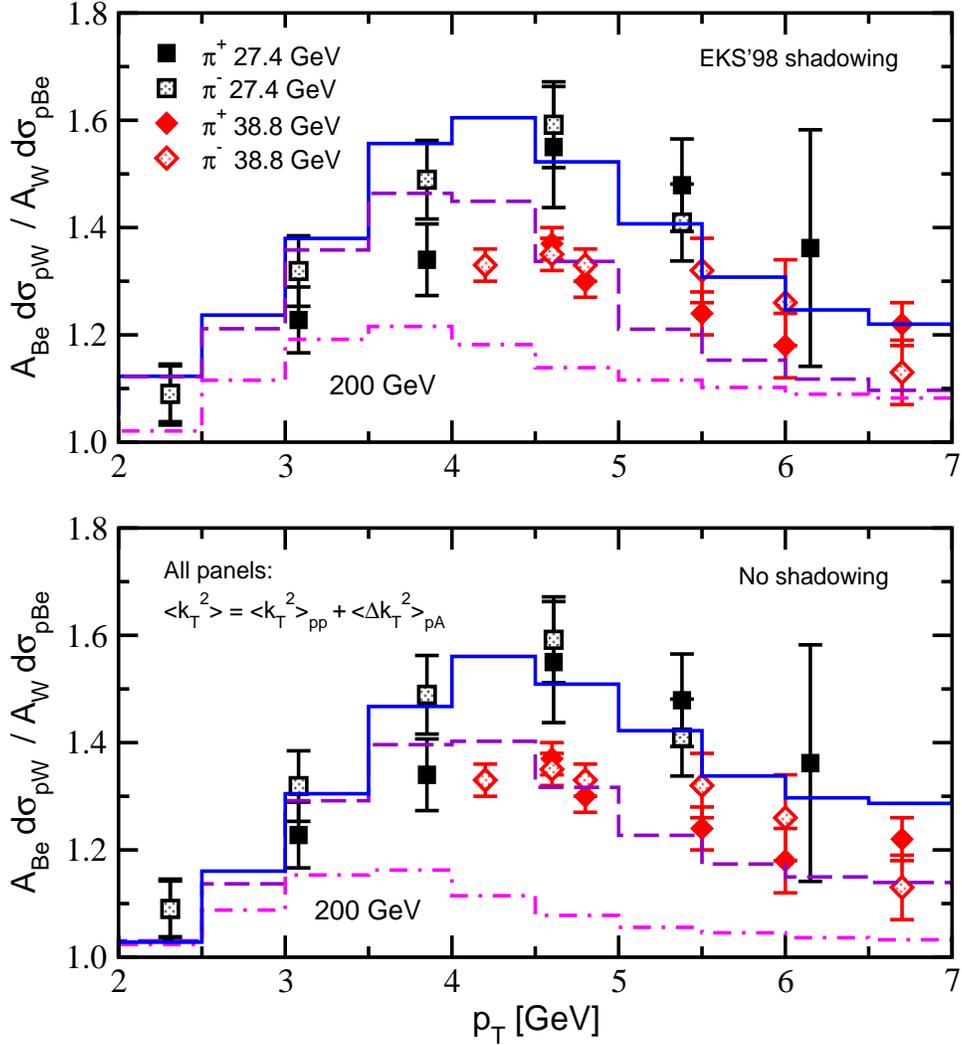} 
\vspace*{0.0cm} 
\caption{The ratio of $A$-scaled  $p+W/p+Be$ data on $\pi^+,\pi^-$  
production at  $\sqrt{s}=27.4, 38.8$~GeV from~{\protect \cite{Cronin:zm}}. 
Calculations are for $\half (\pi^+ + \pi^-$):  top/bottom  panel 
include/do not include anti-shadowing/EMC effect as described in the text.
The anticipated $\sqrt{s}=200$~GeV  $p+W/p+Be$ ratio is  shown 
to illustrate the energy dependence of the Cronin effect.} 
\label{cron_wbe} 
\end{figure}

To implement {\em initial state} elastic and radiative  energy loss we 
focus on the large $Q^2 \simeq p_T^2$  partonic subprocess 
$ab \rightarrow cd$,   
$k_a,k_b$ being the initial momenta involved in the 
hard part $d \sigma^{ab \rightarrow cd} / dt$ of Eq.(\ref{hcrossec}).  
If $a,b$  have lost fractions $\epsilon_\alpha,\, 
\alpha=a,b$  of their 
longitudinal  momenta  according to a probability distributions 
$P_\alpha(\epsilon)$,  at  asymptotic   $t= - \infty$   
\beq
\tilde{k}_\alpha = \frac{1}{1-\epsilon_\alpha}  k_\alpha \; , \qquad  
f_{\alpha/p}(x_\alpha,Q^2) \rightarrow \int d \epsilon_\alpha \, 
P_\alpha(\epsilon_\alpha) f_{\alpha/p}\left( \tilde{x}_\alpha= 
\frac{x}{1-\epsilon_\alpha},Q^2  \right) 
\theta( \tilde{x}_\alpha \leq 1)\;\;.  
\eeq{initloss} 
Eq.(\ref{initloss}) provides a simple modification to the 
factorized pQCD hadron production formalism. 
For bremsstrahlung processes $P_\alpha(\epsilon)$ are sensitive  
to multiple gluon emission~\cite{Gyulassy:2001nm}. 
For the simpler case of mean energy loss  $P(\epsilon) =  
\delta(\epsilon  - \langle \Delta k_0 \rangle / k_0 )$. More 
specifically, for the small elastic longitudinal shift that 
we consider here  
\beq
P_\alpha(\epsilon) =  \delta\left( \epsilon  - 
\frac{\mu^2\chi_\alpha \xi}  {2 k_{\alpha\; \parallel}^2 }  \right),  
\qquad    f_{\alpha/p}(x_\alpha,Q^2) \rightarrow  
f_{\alpha/p} \left( x_\alpha + \frac{\mu^2 \chi_\alpha \xi }{x_\alpha}
\, \frac{2}{s} , Q^2 \right)  \theta( \tilde{x}_\alpha \leq 1)\;\;.
\eeq{elsh}
The observable effect of Eqs.(\ref{initloss},\ref{elsh}) 
can be very different for valence quarks, sea quarks, and gluons  
due to the different $x$-dependence of the PDFs.

Two fits to the $\sqrt{s}=27.4, \, 38.8$~GeV  $p+W/p+Be$ data
are given in  Fig.~\ref{cron_wbe}. In the first case we use the 
EKS'98 shadowing function $S_{\alpha/A}(x_\alpha,Q^2)$ to model  
the antishadowing/EMC modification of the nuclear target. 
The changing quark/gluon projectile composition, 
reflected in  its $\sqrt{s}$-dependent average color charge,  
enters the opacity through the elastic scattering cross section:
\beq
\chi(\sqrt{s}) = \sigma_{el} \rho  \langle L \rangle   
 \propto   \left(  C_F F^h_q(\sqrt{s},\bar{Q}) + 
  C_A F^h_g(\sqrt{s},\bar{Q})   \right)    \bar{C_T} \;\;,
\eeq{opacch}   
i.e. in a way analogous to Eq.(\ref{ktpp}). 
In Eq.(\ref{opacch})  $\bar{C_T}$ is the mean color charge of the 
target.  The variation of the opacity with the center of mass energy 
can thus be inferred from  Table~\ref{table} and is given by 
$\chi(\sqrt{s}) = \Delta(\sqrt{s})\cdot \chi(27.4\,{\rm GeV})$ 
for fixed  $\bar{C_T}$.  While quark antishadowing/EMC effect are  
constrained by data, the modifications to the gluon PDFs are quite 
uncertain. In the second calculation we do not include nuclear 
shadowing. Instead,  in Eq.(\ref{opacch}) we also take 
a changing average color charge for the target: 
$C_T(\sqrt{s})  = \left( \, C_F F^h_q(\sqrt{s},\bar{Q}) +   
C_A F^h_g(\sqrt{s},\bar{Q}) \, \right)$. In this case the variation 
of the effective opacity with $\sqrt{s}$  is given by 
$\chi(\sqrt{s}) = 
\Delta^2(\sqrt{s})\cdot \chi(27.4 \, {\rm GeV})$.    
Both calculations include initial state mean elastic energy 
loss. With the definition of $\langle L \rangle$ given in 
Eq.(\ref{Leff}) the transport coefficients of cold nuclear matter 
are fixed to  $\mu^2/\lambda_q \approx 0.06\; {\rm GeV}^2/{\rm fm}, \,  
\mu^2/\lambda_g = (C_A/C_F)\, \mu^2/\lambda_q  \approx 
0.14\; {\rm GeV}^2/{\rm fm} $.  The modest $\sim 20\%$ increase relative 
to previous estimates  ($\mu^2/\lambda_q \simeq 0.05\; 
{\rm GeV}^2/{\rm fm}$~\cite{Vitev:2002pf}) compensates  
for the longitudinal backward momentum shift, Eq.(\ref{parshift}), 
and provides comparable reproduction of existing low  energy $p+A$ 
results. This can be seen by comparing the top panel of 
Fig.~\ref{cron_wbe} to Fig.~1 from Ref.~\cite{Vitev:2002pf}.  
The bottom panel of  Fig.~\ref{cron_wbe} seems to show better 
agreement between data  
and theory in terms of improved reproduction of the Cronin peak and 
separation of the  nuclear  modification ratios versus $\sqrt{s}$ 
at high $p_T$. We continue to explore both possibilities but 
the calculations without strong antishadowing/EMC effect 
are also expected to give better results at RHIC energies.

The systematic reduction of the Cronin enhancement/suppression 
with increasing  $\sqrt{s}$ has its natural explanation in the 
picture of multiple elastic scattering on a nuclear 
target~\cite{Gyulassy:2002yv} (see Fig.~\ref{cron_wbe}). 
With the opacity $\chi$ changing only slightly with $\sqrt{s}$ the 
dominant effect becomes the significant 
hardening of the  hadronic spectra as predicted by perturbative QCD. 
This reduces the observable consequences of otherwise identical 
transverse momentum  kicks.

We have also checked that the flavor dependence of the Cronin effect  
- the larger enhancement/suppression for kaons (and also protons) 
-  cannot be reproduced within the accuracy of the calculation.  
Studies have so far been focused on 
the baryon enhancement puzzle~\cite{Vitev:2001zn} but the results are 
inconclusive. We note that improved perturbative techniques are 
necessary to address the  Cronin effect for massive hadrons and extend 
the calculations to the $p_T \lton 2$~GeV region.

\section{Impact parameter and rapidity dependence 
of the Cronin effect} 
\label{sec_disc}

\begin{figure}[t!] 
\hspace*{0.cm }\epsfig{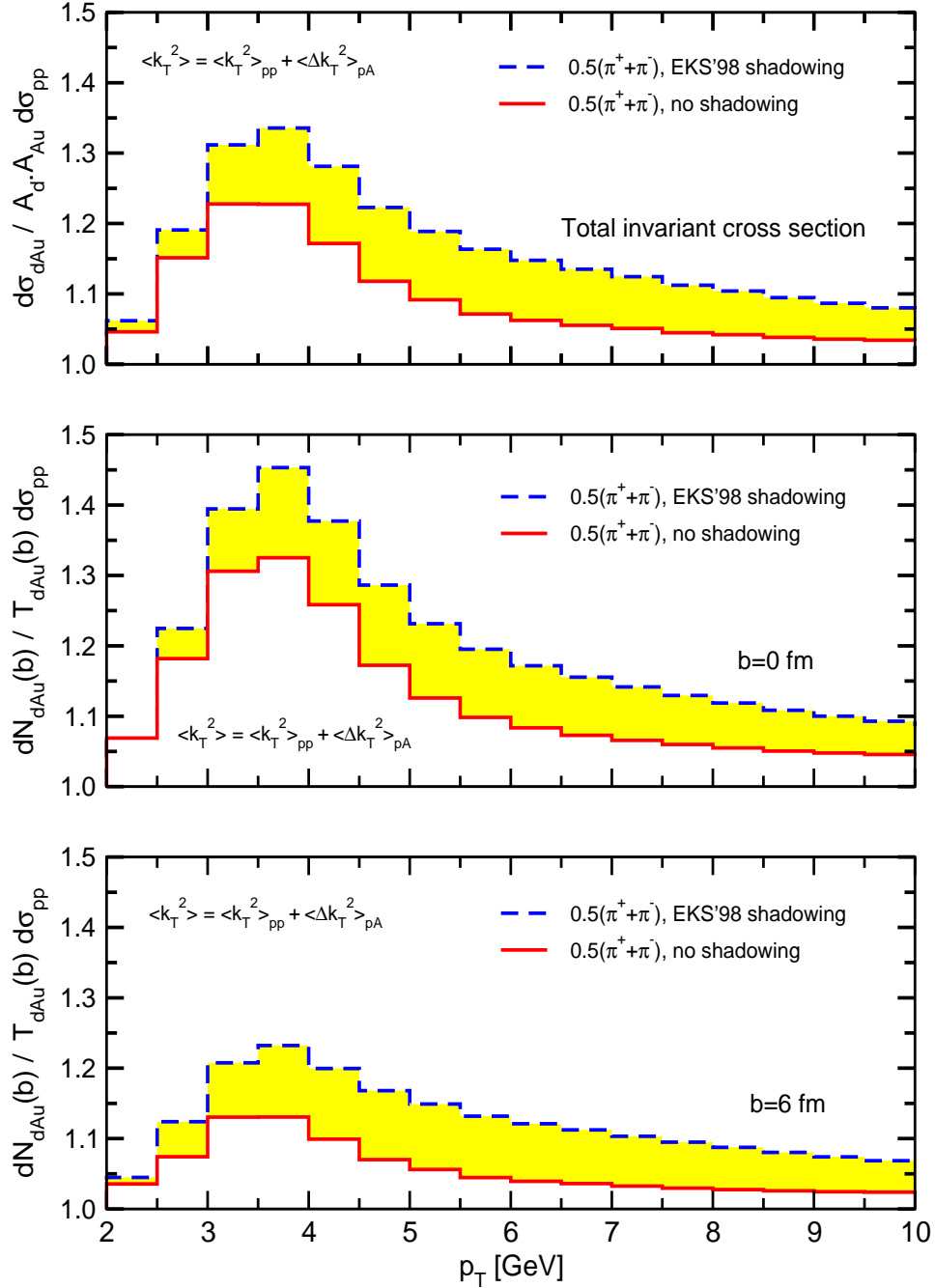} 
\caption{The  Cronin effect,  represented by the 
nuclear modification ratio $R_{dAu}(p_T)$,  for 
$\half(\pi^+ + \pi^-)$ is shown for the total invariant cross 
section and impact parameters $b=0,6$~fm.} 
\label{cronindAu} 
\end{figure}

The $\sqrt{s} = 200$~GeV results for the Cronin effect in $d+Au$ 
reactions at midrapidity are shown in Fig.~\ref{cronindAu}. 
At $y=0$ for the corresponding  $2\times 10^{-2}  \leq x_T 
\leq 10^{-1}$ values  the  shadowing  region of $S_{a/A}(x,Q^2)$ is 
never reached. However, the relative contribution of  antishadowing 
is seen to be large for $p_T \geq 3$~GeV. We find that without significant  
$x$-dependent  modification of  the  nuclear target (i.e. the case 
that gives  better agreement with existing data 
from Fig.~\ref{cron_wbe}) the  
Cronin enhancement reaches  1.2  with peak position 
($R_{dAu \; \max}$) in the  
$p_T \simeq 3-4$~GeV range, a 30\% drop relative to the 
strong  gluon antishadowing scenario. In going from central  
($b=0$~fm) to  peripheral ($b=6$~fm) reactions the magnitude of the 
Cronin effect is reduced  by  a  factor $\sim 2.5$. The main
uncertainty in computing $R_{dA}(p_T)$ for peripheral reactions comes 
from the large deuteron radius, leading to a distribution in 
impact parameter of its partonic constituents. 
The corresponding spread  in opacity $\chi = \sqrt{R_A^2-b^2}/\lambda $  
grows with  increasing $b$ and using the average value 
$\chi(b)$ in  Eqs.(\ref{ropit},\ref{parshift}) may not be a 
good approximation.

\begin{figure}[t!] 
\hspace*{0.cm }\epsfig{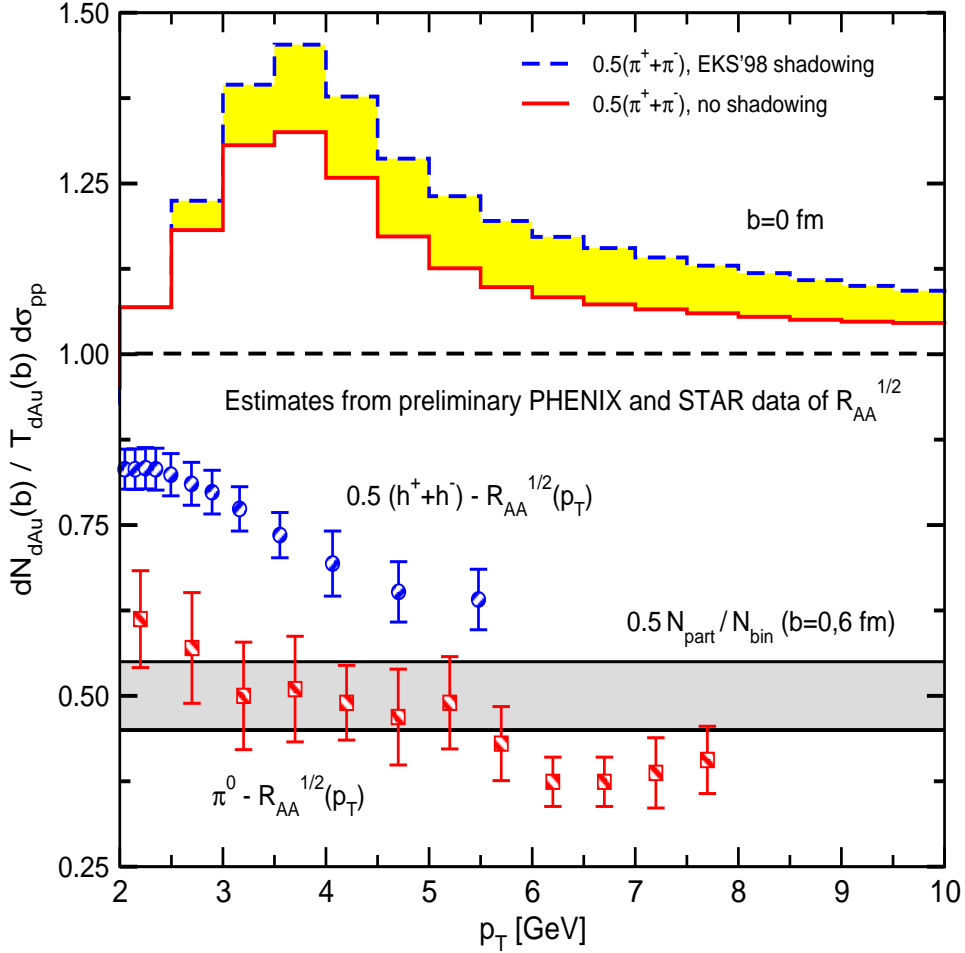} 
\caption{Illustration of the difference between the computed small 
$\sim 5-10\%$  Cronin enhancement at $p_T\geq 5$~GeV in central 
reactions and the argued~\protect\cite{Gallmeister:2002us} factor 
of $30-50\%$  suppression.  The estimated $\sqrt{R_{AA}(p_T)}$ for 
$\half(h^+ + h^-)$ and $\pi^0$ from preliminary STAR and PHENIX 
data~\cite{Adcox:2001jp}  and the $N_{part}/2$ scaling for $b=0,6$~fm
are shown.}  
\label{diffdAu} 
\end{figure}

Midrapidity measurements accessible to all experiments, especially 
for central reactions, are  the prime candidate that can distinguish 
between different hadron production scenarios. Glauber 
model calculations~\cite{Glauber:1970jm} that account for the large 
deuteron radius show a factor $\sim 2$  difference between  
$N_{bin}$ and $N_{part}/2$ (not $N_{part}$!), the two possible  
scalings  relative to $p+p$, over a wide range of impact parameters. 
Participant scaling at moderate and high  $p_T$  
(up to $p_T=8-10$~GeV as discussed in~\cite{Gallmeister:2002us}) 
will correspond to $R_{dAu}(p_T) \simeq 0.5$  as shown in Fig.~\ref{diffdAu}.
The gluon saturation model suggests $R_{dAu}(p_T)= 
\sqrt{R_{AuAu}(p_T)}$, i.e. it will in fact lead to the same conclusion  
since the suppression in $Au + Au$ reactions is 3-5 fold. This is 
demonstrated in Fig.~\ref{diffdAu} through the estimated 
$\sqrt{R_{AuAu}(p_T)}$ for inclusive charged hadrons and neutral pions  
from preliminary STAR and PHENIX  data~\cite{Adcox:2001jp}. 
We emphasize that the $p_T$ dependence of the nuclear modification 
ratio in $d+Au$ is different as well. 
Final state 
hadron absorption mechanisms~\cite{Gallmeister:2002us} are also 
expected  to result in suppression in $d+Au$, albeit smaller 
and difficult to quantify analytically. In contrast, the calculations 
in  Figs.~\ref{cronindAu} and~\ref{diffdAu} show small enhancement 
in the specified $p_T$ range even with initial state elastic energy 
loss. Predictions that have quantitatively addressed the $y=0$, 
$p_T \geq 2$~GeV Cronin  effect at  RHIC from the standpoint of 
perturbative QCD and  multiple elastic scatterings give similar
results~\cite{Vitev:2002pf,Accardi:2002ik}. 
Experimental measurements in $d+Au$ will enable a critical test to 
differentiate between competing interpretations of $Au+Au$ data.

\begin{figure}[t!] 
\hspace*{0.cm }\epsfig{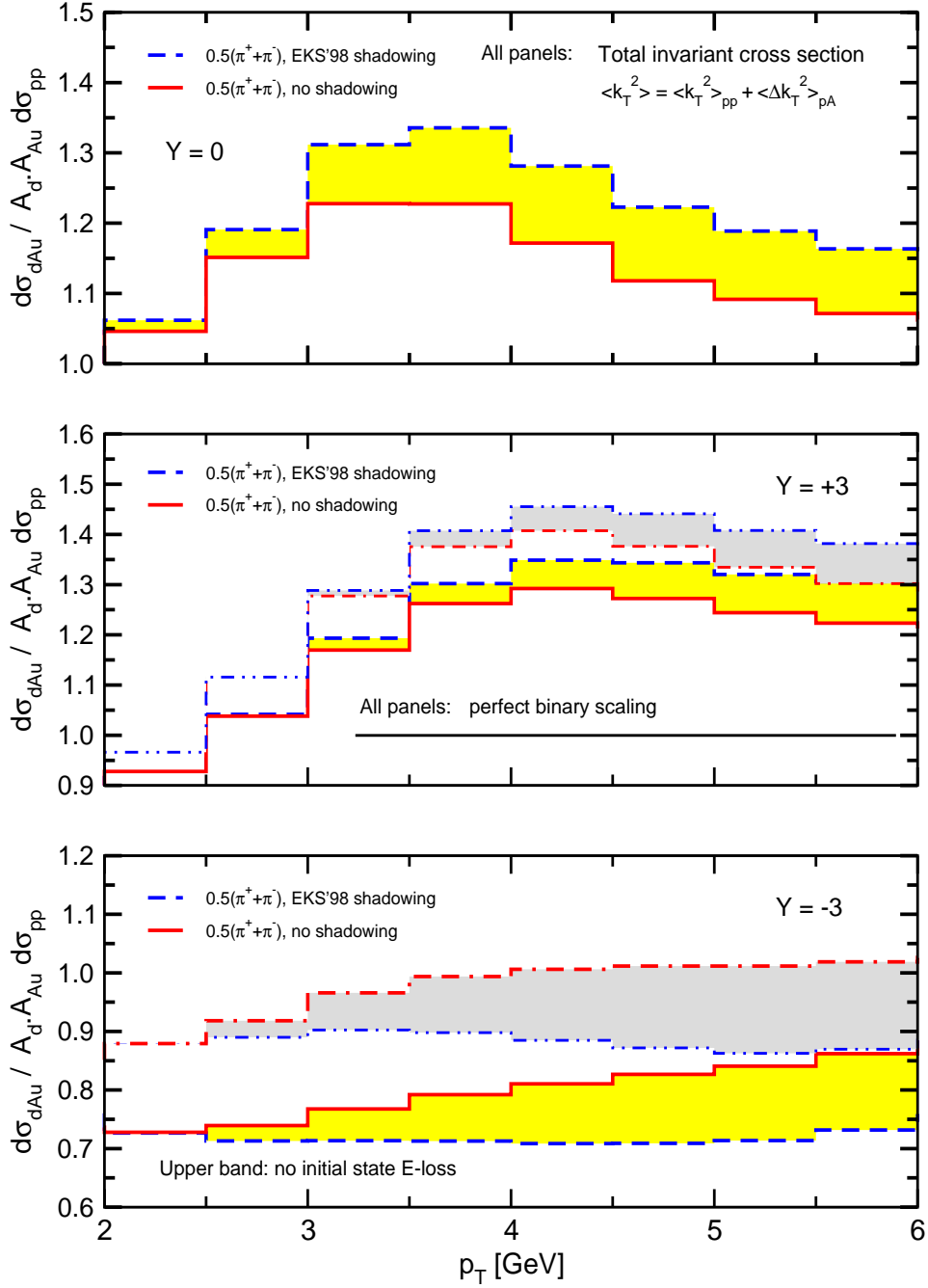} 
\vspace*{.2cm} 
\caption{ Rapidity dependence of the Cronin effect in $d+Au$ 
reactions  at $\sqrt{s}_{NN}=200$~GeV with and without 
antishadowing/EMC effect. The result of switching off elastic 
energy loss is also shown via the upper bands for $y=\pm 3$. }
\label{cron_rap} 
\end{figure}

Fig.~\ref{cron_rap} shows the predicted Cronin effect at forward 
($y=+3$) and  backward  ($y=-3$) rapidities near the edge of BRAHMS 
acceptance. In the following  calculations the direction of the deuteron 
beam  is chosen to be positive. At forward rapidity, which probes the 
small-$x$ region of the nucleus, the effect of shadowing is very  
small. Similarly, turning off elastic energy loss does not 
significantly change the shape and magnitude of $R_{dAu}(p_T)$.
Relative to $y=0$,  both the peak and intercept ($R_{dAu}=1$) 
positions are shifted to slightly higher transverse momenta.   
The most distinct prediction at $y=+3$ is the significantly flatter 
Cronin enhancement region that extends to high $p_T$. At 
forward/backward rapidities hadron spectra are much steeper than 
at $y=0$. This explains the markedly broader range  where  
the effect of transverse momentum  kicks is observed.

At backward  $y=-3$  rapidity, shown at the bottom panel 
of Fig.~\ref{cron_rap}, there is no Cronin enhancement since the 
partons from the nucleus do not scatter multiply on the 
deuteron. However, in the $p_T$ range shown data probes the large 
$x \gton 0.4 $ nuclear modification $S_{a/A}(x, Q^2)$ and may provide 
complementary information about the EMC effect. 
The backward rapidity region  is also  more sensitive to initial state 
energy loss.  Switching it off  changes $R_{dAu}(p_T)$ from  
moderate $\sim 25 \% $  suppression to an almost perfect binary 
collision scaling  ($R_{dAu}(p_T) \simeq 1$) for $p_T \geq 3$~GeV.

\section{Conclusions}

In this letter we have derived a technique of incorporating initial 
state parton energy loss in perturbative calculations. 
Midrapidity moderate- and high-$p_T$ hadron spectra show little
or no sensitivity to this effect since the longitudinal momentum 
backward shifts can be compensated by a small increase in the estimated 
cold nuclear transport coefficients to obtain an equally good description 
of existing $p+A$ data. Detecting initial state energy loss  
may be easier in the small-$x$ region of the deuteron where the steepening 
of the parton distribution functions would tend to amplify the effect.

In this letter  we have given predictions for the centrality and 
rapidity dependence, see Figs.~\ref{cronindAu} and~\ref{cron_rap}, 
of the moderate and high transverse  momentum  ($ p_T \geq  2$~GeV) Cronin 
effect in $d+Au$ reactions at RHIC. They follow the systematic approach 
to the computation of the nuclear modification factors outlined in 
Ref.~\cite{Vitev:2002pf}.  At midrapidity we find a small $\sim 20-30 \%$  
Cronin enhancement that decreases in going from central to peripheral 
reactions and at high $p_T$. Forward and backward ($y=\pm 3$) 
rapidity regions are predicted to have markedly different  $R_{dA}(p_T)$.     
Upcoming  central $d+Au$ data will provide a decisive test to 
distinguish between suppression and enhancement dynamical 
models~\cite{Vitev:2002pf,Gallmeister:2002us,Accardi:2002ik}, 
see Fig.~\ref{diffdAu}, 
of initial state effects in nuclear collisions.

\acknowledgments

Many illuminating discussions with  M.~Gyulassy, J.W.~Qiu, and 
J.P.~Vary are gratefully  acknowledged. I would like to thank 
J.~Klay and D.~d'Enterria for clarifying comments on the 
preliminary STAR and PHENIX data. This work is supported  by the 
United States Department of Energy under Grant No. DE-FG02-87ER40371.

\end{document}